Machine Learning Based Radiomics for Glial Tumor Classification and Comparison with Volumetric Analysis

Sevcan Turk[1,6], Kaya Oguz[2], Mehmet Orman[3,] Emre Caliskan[4], Yesim Ertan[5], Erkin Ozgiray[4], Taner Akalin[5], Ashok Srinivasan[1], Omer Kitis[6]

1. University of Michigan, Radiology Department, Ann Arbor, MI, USA
2. Izmir University of Economics, Department of Computer Engineering, Izmir, TR
3. Ege University School of Medicine, Biostatistics Department, Izmir, TR
4. Ege University School of Medicine, Neurosurgery Department, Izmir, TR
5. Ege University School of Medicine, Pathology Department, Izmir, TR
6. Ege University School of Medicine, Radiology Department, Izmir, TR

Corresponding Author : Sevcan Turk, MD

email: sevcant@med.umich.edu

Address: 1500 East Medical Center Drive, UH B2A29, 48109,

Ann Arbor, USA

Original Work Place; Ege University Hospital, 35100, Bornova, Izmir, Turkey

Manuscript Type: Original Research

Funding: Sevcan Turk got RSNA 2018 Student Travel Award.

The authors have no relevant financial or non-financial interests to disclose.

Conflict of interest: No conflict of interest.

Original Research Article

Summary Statement:

Radiomics and machine learning are useful techniques to extract information that is not apparent in routine interpretations. Application of machine learning methods to MRI features can be used to classify brain tumors more readily in clinical settings.

Key Results:

- When we compared the volume ratios between groups, there was statistically significant difference between grade IV and grade II-III glial tumors.
- Edema and tumor necrosis volume ratios for grade IV glial tumors were higher than that of grade II and III.
- Volumetric ratio analysis could not distinguish grade II and III tumors successfully, however, SVM and ANN correctly classified each group with accuracies up to 99% and 96%.

List of Abbreviations:
SVM: Support Vector Machine
ANN: Artificial Neural Network
AUC: Area Under the Curve
ROC: Receiver Operating Characteristic Curve
WHO: World Health Organization
AI: Artificial Intelligence
FLAIR: Fluid attenuated inversion recovery
ADC: Apparent diffusion coefficient

## Abstract

### Purpose;

The purpose of this study is to classify glial tumors into grade II, III and IV categories noninvasively by application of machine learning to multi-modal MRI features in comparison with volumetric analysis.

### Methods;

We retrospectively studied 57 glioma patients with pre and postcontrast T1 weighted, T2 weighted, FLAIR images, and ADC maps   acquired on a 3T MRI. The tumors were segmented into enhancing and nonenhancing portions, tumor necrosis, cyst and edema using semiautomated segmentation of ITK-SNAP open source tool. We measured total tumor volume, enhancing-nonenhancing tumor, edema,  necrosis volume and the ratios to the total tumor volume. Training of a support vector machine (SVM) classifier and artificial neural network (ANN) was performed with labeled data designed to answer the question of interest. Specificity, sensitivity, and AUC of the predictions were computed by means of ROC analysis. Differences in continuous measures between groups were assessed by using Kruskall Wallis, with post hoc Dunn correction for multiple comparisons.

### Results;

When we compared the volume ratios between groups, there was statistically significant differen
ce between grade IV and grade II-
III glial tumors. Edema and tumor necrosis volume ratios for grade IV glial tumors were higher th
an that of grade II and III. Volumetric ratio analysis could not distinguish grade II and III tumors
successfully. However, SVM and ANN
correctly classified each group with accuracies up to 98% and 96%.

**Conclusion;**

Application of machine learning methods to MRI features can be used to classify brain
tumors noninvasively and more readily in clinical settings.

**Background:**

MRI plays an essential role in the management of neuro-oncology patients, but the
interpretation of the images remains subjective and for the most part non-quantitative. Artificial
intelligence based systems and radiomics are becoming essential methods enabling analysis of
extracted quantitative features from radiographic images. Previous studies have shown the
accuracy and reproducibility of machine learning algorithms to demonstrate tumor shape,
intensity and texture characteristics from MRI (1,2). Volumetric assessment of the tumors is also
shown to be useful to predict genetic mutation and survival rates (3,4). There is a significant
correlation between the amount of necrosis within a tumor and the tumor grade, and intra-
tumoral heterogeneity associated with higher tumor grade, treatment resistance, and poor
prognosis. The grading of gliomas has a clinical impact on therapeutic decision-making,
monitoring, and prognostic prediction. Since pathologic biopsy evaluations depend on a slice of
tissue rather than the entire tumor itself, analysing the tumor components in a whole-volumetric
fashion may provide further information about its heterogeneity. We hypothesized that
heterogeneity of the tumor components, surface characteristics, shape, and intensity would add
value to the success of volumetric data derived from conventional MRI studies.

**Purpose;**

The purpose of this study is to classify gliomas into grade II, III and IV categories noninvasively
by application of machine learning algorithms and to show the superiority of artificial intelligence
(AI) over conventional volumetric analysis.

**Methods;**

Institutional review board approval was taken for this retrospective research. Informed consent
was waived. A total of 57 glioma patients scanned on a 3 Tesla MRI (Verio, Siemens, Erlangen,
Germany)  were reviewed in three groups: 18 WHO grade II, 14 grade III, 25 grade IV, who
presented at our institution between 2006-2017. Each sample was graded by two
neuropathologists with more than 15 years experience according to the WHO criteria.
Preoperative MR images were retrospectively evaluated in a blinded manner. Pre- and post-
contrast axial T1W (TR: 450 ms, TE: 9.4 ms, slice thickness: 5mm, NEX:1), axial T2W (TR:
3010 ms, TE: 117 ms, slice thickness: 5 mm, NEX: 1), coronal FLAIR (TR: 9000 ms, TE:92 ms,
TI:2500 ms, slice thickness: 4 mm, NEX:1) and axial ADC maps (DWI TR:8100 ms, TE:81 ms,

slice thickness: 5 mm, NEX: 2, b: (0)-(1000) ) were included in the analysis. Subtraction maps were created by aligning the slices of pre-contrast and post-contrast T1 weighted images using multi-modal registration that uses Mattes mutual information metric and One-plus-One evolutionary optimizer. Then, the pre-contrast images were subtracted from the post-contrast images and voxels with a positive value were retained in the subtraction map for analyses. A 4[th] year radiology resident segmented the tumors slice by slice into 5 labels blindly under the supervision of a neuroradiologist with 18 years experience: Enhancing (label-2) and non-enhancing solid components (label-3), necrosis (label-5), cysts (label-4), and perilesional edema (label-1) by semiautomated and manual segmentation using open-source software (ITK-SNAP ver. 3.6)(Figure 1). Volumetric calculations of each of the five components and their ratios to the total tumor volume were analyzed with ITK-SNAP and SPSS.

To quantify the spatial heterogeneity of the tumor components amongst the three groups of Grade II, Grade III and Grade IV (G II, G III and G IV, respectively), we used the texture and histogram features that are detailed as follows (Figure 2). For all regions of interest that have been labeled, a feature vector $V_1$ is created that consists of the histogram of these regions using 10 bins, minimum, maximum and mean values, and the entropy value. The normalization of histograms was done by using 10 bins for the intensity values of the segmented area, rather than the same intervals for all areas, therefore making them reflect the true distribution of their intensity values. The entropy value is the Shannon Entropy defined as $E = -\sum p \log 2p$ where $p$ is the probability. Entropy shows how random the texture pixels are, giving an insight to the heterogeneity of the ROI. $V_1$ has a length of 14. The next set of features, $V_2$, are the concatenation of the same histogram, minimum, maximum, mean and entropy values for each of the five regions of interest. If there is no such label in a volume, then those values are left as zeros. This information is also important since the learning algorithm can notice the absence of some labels. $V_2$ has a length of $14 \times 5 = 70$. The final set of features, $V_3$, only use label-2 and label-5 using the same set of histogram, minimum, maximum, mean and entropy values, and has a length of $14 \times 2 = 28$.

Shape analyses of each component of the tumor were also used to train both SVM (support vector machine) and ANN (artificial neural network). These features were extracted from the masks that are created for each label. Each mask was parted using a standard connected component approach and then the following features were extracted. Solidity is calculated as the ratio of the area to the convex area of the mask. The other features are calculated with respect to the ellipse that confines the area. Eccentricity is the ratio of the distance between the foci of the ellipse to its major axis length. The ratio of the major axis to the minor is set as the third feature. The perimeter of the ellipse is calculated by Ramajunan approximation. The ratio of this perimeter to the perimeter of the area is set as the fourth feature.

Since SVM is a binary classifier, one-against-all approach is used to train three classifiers for GII versus G III and G IV, G III versus G II and G IV, and G IV versus G II and G III, respectively. All classifiers have been trained by using two different kernels, linear and radial basis. The ANN classifier consisted of 20 hidden neurons which were trained by conjugate gradient backpropagation algorithm (Figure 3). Classification is done by minimizing the cross-entropy. For both SVM and ANN, 80% of the data was allocated only for training, 10% of the data was allocated for validation, and the final 10% was used for testing. Training-testing and validation data sets were selected randomly.

The ANN paradigm was trained 100 times and the mean and the best test results were then reported. The results of the ANN were used as the second classifier and then compared to those obtained by SVM. The ANN classifier afforded simultaneous comparisons between the three groups whereas SVM could perform comparisons separately between only two groups at a time. This was considered one of the advantages of ANN over SVM.

Statistical analysis was performed using IBM SPSS version 25.0 software. Specificity, sensitivity, and AUC of the predictions were computed utilizing ROC analysis. Differences in continuous measures between groups were assessed by using Kruskal-Wallis test, with post hoc Dunn correction for multiple comparisons. For all statistical tests, $p<0.05$ was considered statistically significant.

**Results:** There were 20 female and 37 male patients in the study. The mean age was 37 (SD 14) years for grade II, 34 (SD 15) years for grade III and 53 (SD 16) years for grade IV glial tumors; the overall mean age was 43 (SD 17) years.

Comparisons of volume ratios between groups showed a statistically significant difference between grade IV and other grades (II-III) glial tumors in all tumor component ratios separately ($p= 0.02$). There was no significant statistical difference between the total tumor volumes amongst groups ($p=0.13$). Edema, necrosis and enhancing solid tumor part ratios for grade IV glial tumors were higher than that of grade II and III ($p<0.001$). The median edema volume ratio for grade IV tumors was 43% with 0.83/80.5 min/max values while it was 2.75% with 43% with 0/81.1 min/max values in the grade III and 0% with 0/20.9 min/max values in the grade II tumors (Table 1). Median necrosis volume ratio for grade IV glial tumors was 1.3% with 0/17.6 min/max values whereas it was < 1 % for both grade II and III glial tumors with 0/8.2 and 0/2.5 min/max values (Table 1). Median solid enhancing volume ratio for grade IV group was 41% with 12.9/85.9 min/max values where it was 7% with 0/34 min/max value for grade III and 0.16% with 0/21 min/max values for grade II tumors. Median non-enhancing tumor volume ratio was statistically higher in grade II and III tumors than grade IV ($p<0.001$)(Table 1). Median solid non-enhancing tumor volume was 96.9% with 0/100 min/max values for grade II, 86.6 with % with 0/100 min/max values for grade III and 7.8% with 0/48.7 min/max values for grade IV glial tumors. The cystic volume ratio did not have a statistically significant difference between groups ($p=0.44$). However, this ratio tended to be higher with lower tumor grades. There were no statistically different volume ratios between grade II and III glial tumors for each tumor part.

ANN classified grade II and III tumor groups up to 95% accuracy by using only pre-contrast T1 images texture analysis (mean 79%). The accuracy with post-contrast T1 image texture analysis was 85% (mean 74%) and 70% (mean 63%) with shape feature analysis. SVM accuracy to differentiate grade 2 and 3 was 65% with the linear kernel for both pre-post contrast T1 and subtraction images (Table 2).

ANN was overall superior to SVM to categorize tumor grade with histogram and texture analysis. Histogram analysis was superior to texture and shape analysis to make an accurate classification. ANN classified grade II and IV tumors up to 98% accuracy using ADC maps histogram analysis (mean 95%, mode 98%) (Table 2). Pre-post contrast T1 and T2 images had similar results with ANN histogram analysis which was superior to SVM. On the other hand, SVM classified grade II and IV or grade III and IV tumors with 99 % accuracy (Figure 4). ADC histogram analysis with ANN separated each group up to 89% accuracy (mean 85%) and differentiated grade III and IV with 88 % mean accuracy (best 92%). Histogram analysis of FLAIR images with ANN classified grade II and III correctly up to 96% (mean 69%) (Figure 6)(Table 2).

Volumetric ratio analysis could not distinguish grade II and III tumors. However, AI correctly classified each group with accuracies up to 96%.

**Discussion:** Grading of glial tumors is one the basic classicification which correlates well with patient's treatment and outcome. Noninvasive classification reduces biopsy-surgery related complications especially for grade II tumors which may not require immediate surgical evaluation.

In our study we compared volumetric analysis and radiomics based machine learning algorithms (SVM and ANN) for grade based glial tumor classification. We concluded that artificial intelligence is superior to volumetric analysis to discriminate groups.

Henker et al. searched for glioblastoma volumetric data and overall survival relationship. They segmented the tumor into enhancing, non-enhancing, necrosis, edema components. Tumor necrosis volume related with decreased survival (5).

Lee et al. showed semi-automated segmentation tools' high stability ( ICC >0.8 over 90%) and reliability (6). However, we used both manual and semi-automated segmentation in our study which is the more reliable than automated and semi-automated technics alone. Since manual segmentation is a highly time-consuming process than semi-automated segmentation, we suggest using both together is a more optimal technique. Semi-automated segmentation reported to be expert dependent in a study held by Odland et al. (7).

Cao et al. studied low grade glial tumor and glioblastoma differentiation comparing nine machine learning algorithms and concluded that SVM is the most successful method with an >90% AUC (8). Gates et al. classified grade II, III, IV glial tumors using Random Forest with 89% accuracy. When they added perfusion maps accuracy turned to 96% (9).

Tian et al. used SVM for grading of gliomas and 98% AUC to differentiate high and low grade groups (10). Yang et al. reached 87% accuracy and 97% AUC using SVM to classify grade II, III, IV glial tumors. They also concluded that gray level size zone matrix model is the best SVM classifier when combined with gray level 64 (11). Nakamoto et al. used SVM to differentiate grade 3 and grade 4 gliomas and reached to 93% AUC which was higher than other machine learning methods they compared (12). Alis et al. used ANN to classify gliomas into high and low grade groups and reached 88% accuracy at highest (13).

Gutta et al. showed that convolutional neural networks (CNN) classified grade I, II, III, IV glial tumors automaticly with 87% accuracy (14). Zhuge et al. used CNN to classify glial tumors as low and high grade. They reached 96-87% accuracy and 93% sensitivity with an 96% specificity (15).

Increased volume of necrosis or perilesional edema favors higher tumor grade as well as worse prognosis. Quantitative measurements of these components using volumetric analysis provided discrimination of grade IV and grade I glial tumors from others. However, this tool was not successful to differentiate grade II and grade III which are known as low and high grade gliomas due to the different prognosis and treatment options. Previous volumetric studies did not compare their success with artificial intelligence (5, 7). Our study has shown that texture analysis and machine learning algorithms can provide additional information rather than comparing tumor components as volumetric data. SVM and ANN machine learning tools in this context can provide high accuracy for glial tumor grading. Our study support the available literature reaching up to %96 accuracy to discriminate high and low grade glioma and extends beyond this classification as successfully differentiating even grade II and grade III tumors with up to 98% accuracy.

There are some limitations to the study. This includes retrospective nature of the study and lack of CDKN2A mutation analysis, which is known to be affect the overall patient survival. Also, using coronal 2D FLAIR images created misregistration problems with axial post-contrast T1W images which required more manual segmentation from radiologists.The strength of our study includes the utilization of the same scanner for all studies analyzed; using the same scanner reduces the inherent heterogeneity in acquisition protocols that can arise when including different scanners and increases the accuracy of machine learning algorithm predictions.

**Conclusion**:

Machine learning techniques of SVM and ANN were superior in predicting grades of glial tumors than volumetric evaluations of separate components. Manually segmented volumetric assessment of solid, cystic and necrotic components of the tumors was suboptimal compared to the histogram analyses, texture analyses and shape evaluations based on machine learning techniques. Our study showed the importance and superiority of radiomics information for making classifications more readily in clinical settings.

Image-Based Support Vector Machine for Classifying Glioma. J Magn Reson Imaging. 2019 May;49(5):1263-1274. doi: 10.1002/jmri.26524. Epub 2019 Jan 9. PMID: 30623514.

Tables:

| Table 1 | Grade II Median (min:max) | Grade III Median (min:max) | Grade IV Median (min:max) |
|---|---|---|---|
| edema ratio | 0 (0:20.93) | 2.75 (0:81.14) | 43.47 (0.83:80.54) |
| enhancing solid tumor part ratio | 0.16 (0:21.35) | 7.03 (0:34.01) | 41.18 (12.99:85.97) |
| necrosis ratio | 0 (0:8.29) | 0 (0:2.52) | 1.30 (0:17.63) |
| nonenhancing solid tumor part ratio | 96.91 (0:100) | 86.67 (0:100) | 7.83 (0:48.79) |
| cystic area ratio | 0 (0:100) | 0 (0:12.08) | 0 (0:6.50) |

**Table 1:** Enhancing and non-enhancing tumor part, edema, necrosis, cystic component ratios of grade 2, grade 3 and grade 4 glial tumors.

| Table 2 precontrast T1 | SVM texture | ANN texture mean | ANN texture best | SVM histogram | ANN histogram mean | ANN histogram best |
|---|---|---|---|---|---|---|
| grade II-IV | 99% | 83% | 92% | 61% | 95% | 98% |
| grade III-IV | 99% | 69% | 76% | 73% | 83% | 90% |
| grade II-III | 65% | 79% | 95% | 64% | 76% | 84% |
| all groups | - | 84% | 93% | - | 83% | 88% |

**Table 2:** SVM and ANN trained with texture and histogram analysis AUC values to differentiate groups.

**Figures:**

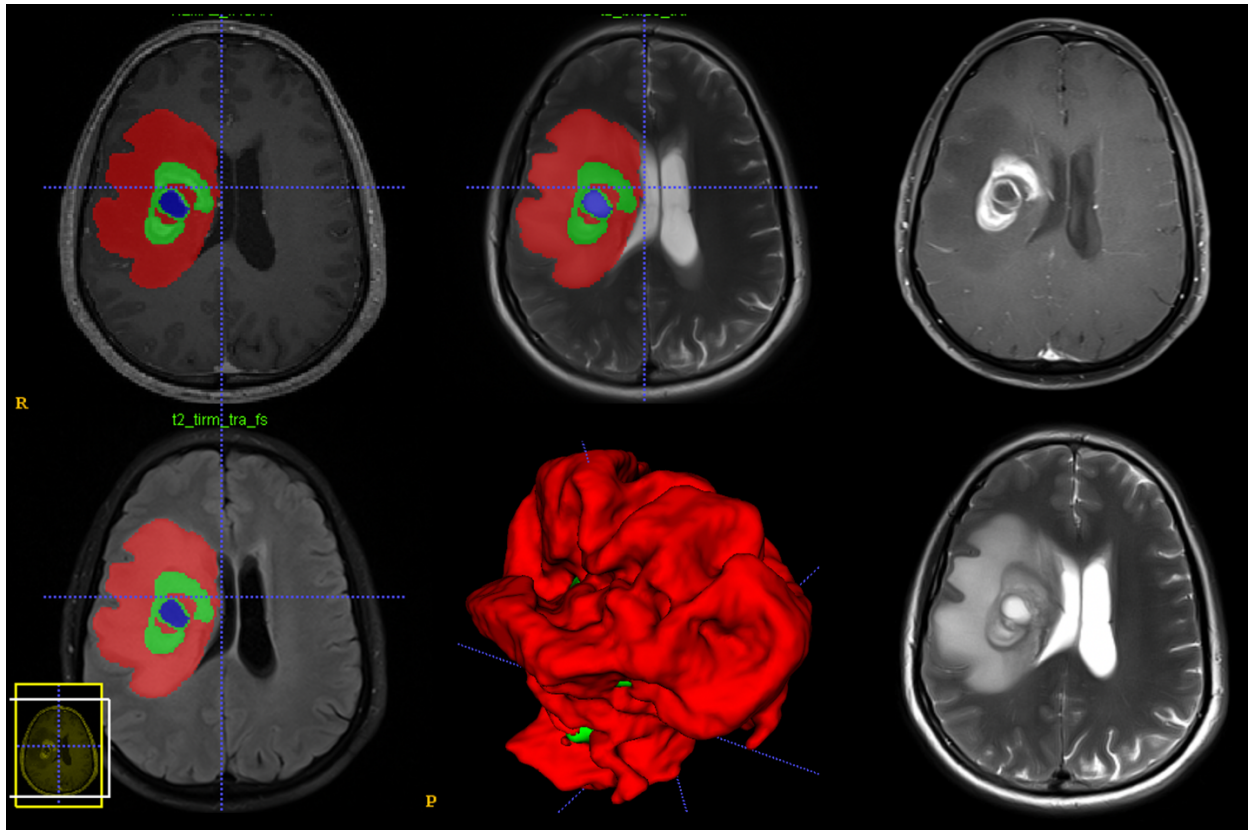

**Figure 1:** Segmentation and 3D image of the tumor; this case is an IDH 1 wild-type

glioblastoma; red area shows edema, green area shows enhancing solid tumor part, blue shows necrosis.

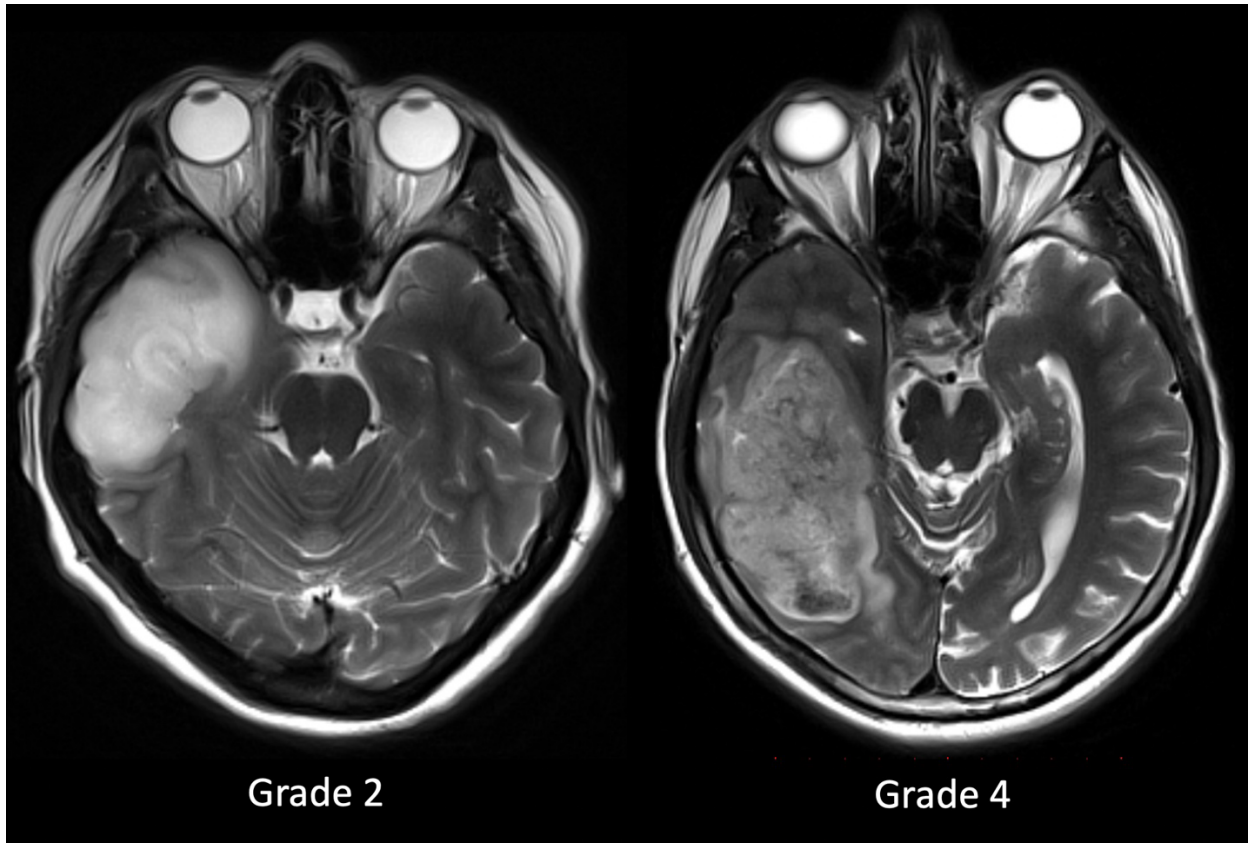

**Figure 2**: Grade II and grade IV glial tumors' T2 weighted images demonstrate different level of visible tumor heterogeneity which increases with tumor grade.

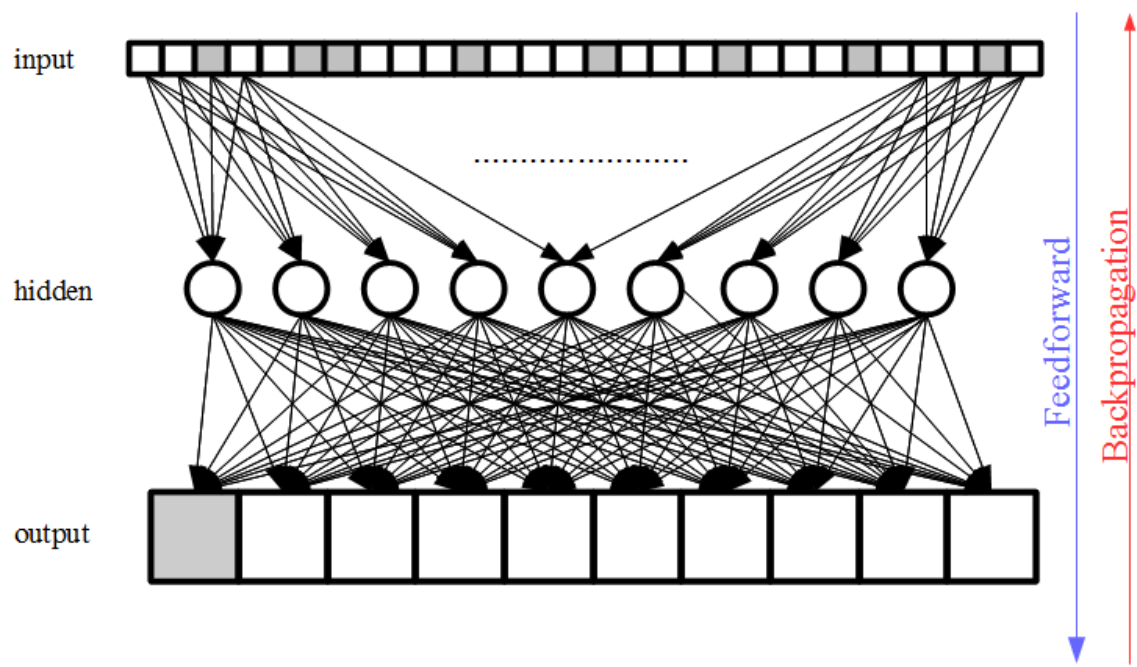

**Figure 3:** Artificial neural network take a set of inputs, integrate data with hidden layers to make predictions.

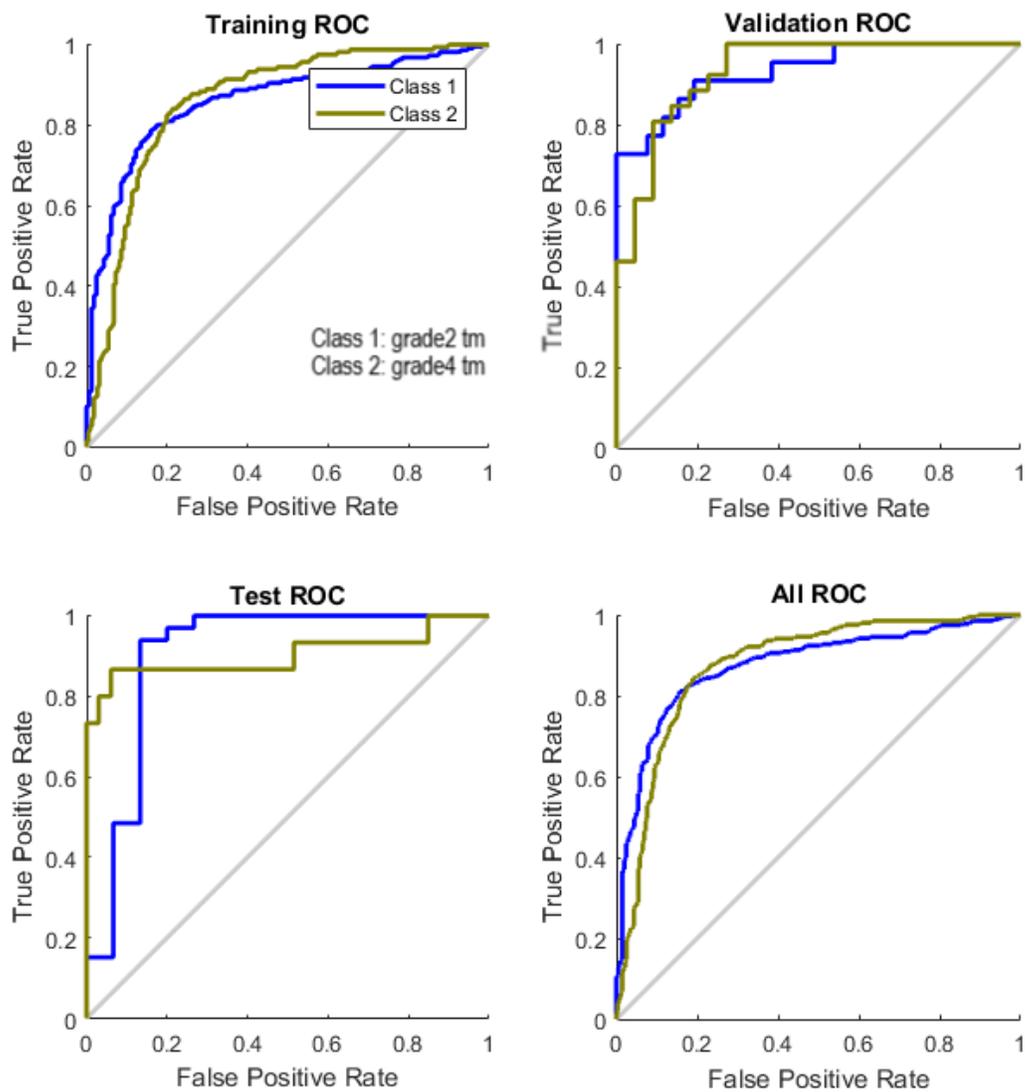

**Figure 4:** Texture analysis comparison of grade II and grade IV glial tumors with SVM which shows accuracies up to 100%.

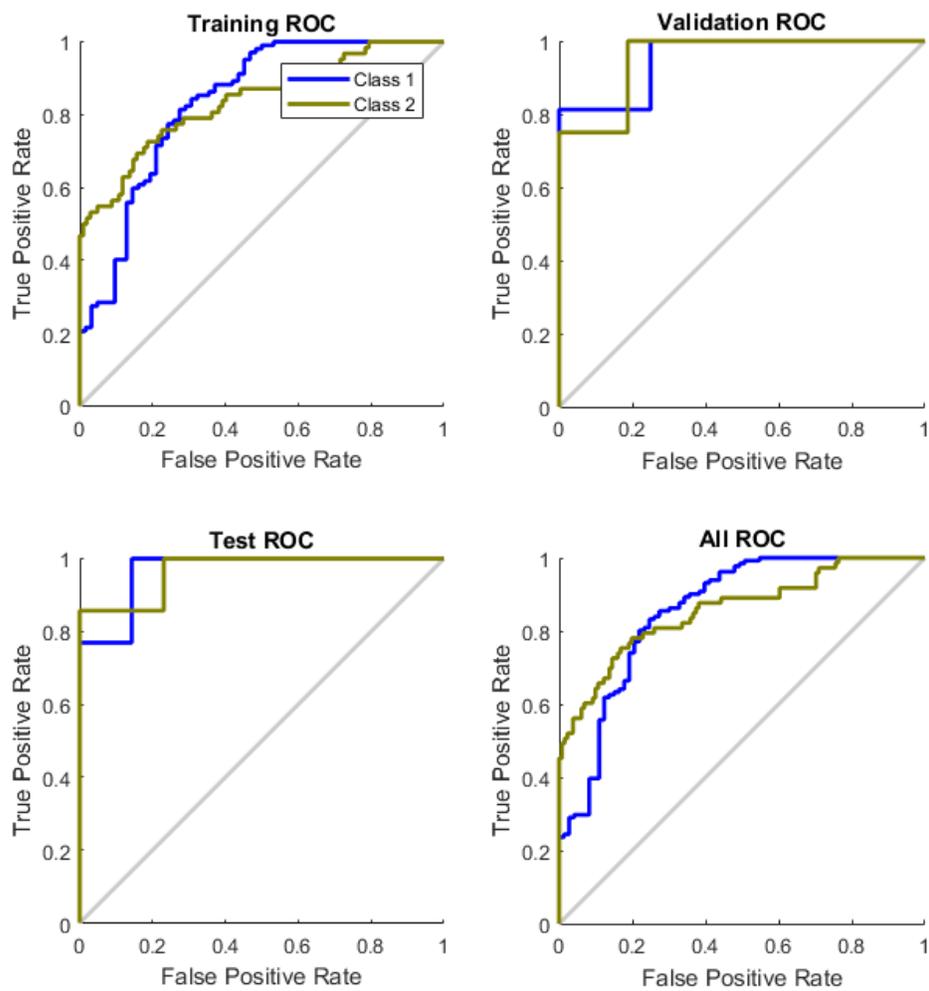

**Figure 5:** ANN classified grade II and III tumor groups up to 95% accuracy by using only pre-contrast T1 weighted images' texture analysis (mean 79%).